\documentstyle[psfig]{l-aa}         

\def\ros{{\sl ROSAT }}
\def\ein{{\sl Einstein}}
\def\asca{{\sl ASCA }}
\def\ginga{{\sl Ginga }}
\def\exo{{\sl EXOSAT}}

\def\it{\sl}
\def\degs{\ifmmode ^{\circ}\else$^{\circ}$\fi}
\def\amin{\ifmmode ^{\prime}\else$^{\prime}$\fi}
\def\asec{\ifmmode ^{\prime\prime}\else$^{\prime\prime}$\fi}

\def\approxlt{\mathrel{\hbox{\rlap{\lower.55ex \hbox {$\sim$}}
        \kern-.3em \raise.4ex \hbox{$<$}}}}
\def\approxgt{\mathrel{\hbox{\rlap{\lower.55ex \hbox {$\sim$}}
        \kern-.3em \raise.4ex \hbox{$>$}}}}

\begin{document}
 
   \thesaurus{03         
              (11.01.2;  
               11.09.1;  
               11.17.2;  
               11.19.1;  
               13.25.2)  
}

   \title{A \ros observation of the warm-absorbed soft X-ray spectrum of NGC 4051}
   \author{Stefanie Komossa, Henner Fink} 

   \offprints{S. Komossa, skomossa@mpe-garching.mpg.de}
 
  \institute{Max-Planck-Institut f\"ur extraterrestrische Physik,
             85740 Garching, Germany\\
       }

   \date{Received 23 July, 1996; accepted }
   \maketitle\markboth{S. Komossa, H. Fink~~~ The soft X-ray spectrum of NGC 4051}{}  

   \begin{abstract}
We present and analyze a pointed \ros PSPC observation of the
Seyfert galaxy NGC 4051. The X-ray spectrum 
consists of a powerlaw modified by absorption edges and an additional 
soft excess during the high-state in source flux. 
Modeling the spectrum in terms of warm absorption yields a large column density of 
the ionized material of  
$\log N_{\rm w}$ = 22.7 and an 
ionization parameter of $\log U$ = 0.4.  
These properties are essentially constant throughout
the observation, whereas the luminosity changes by more than a factor
of 4. The underlying powerlaw is  
in its steepest observed state, with a photon index $\Gamma_{\rm{x}}$ = --2.3. 
Consulting information from optical observations, 
evidence for a separate EUV bump component in NGC 4051 is provided.  
The impact of several parameters 
on the deduced properties 
of the ionized material is critically assessed.
In particular, the influence of dust mixed with the warm gas is explored and shown 
to constrain the history or density of the absorber.            
The absorber-intrinsic optical-UV emission (and absorption) line spectrum is predicted  
and the possibility of a warm absorber origin of one of the observed emission line regions 
in NGC 4051 is investigated. 
Consequences for the narrow-line Seyfert 1 character of NGC 4051 are discussed.
 

   \end{abstract}
 
\section{Introduction}

NGC 4051 is a spiral galaxy of morphological type SAB 
with a redshift of $z$ = 0.0023.  It hosts a  
low-luminosity Seyfert 1 nucleus (Seyfert 1943).  
The object is well known for its rapid X-ray variability which was discovered by
Marshall et al. (1983) with the \ein~observatory. Subsequently, NGC 4051 has been 
extensively studied in the X-ray spectral region with all majour X-ray missions
(e.g. Lawrence et al. 1985, with \exo; Matsuoka et al. 1990, with \ginga;
Pounds et al. 1994, with \ros; Mihara et al. 1994, with \asca).  

The soft X-ray spectrum of NGC 4051 shows several components. 
A black-body-like
soft excess is seen in source high-states (e.g. Pounds et al. 1994, Mihara et al. 1994).
In addition, there is occasional evidence for cold absorption in excess of the Galactic value
(Walter et al. 1994). Around 0.8 keV the spectrum exhibits absorption features,
interpreted 
and modeled 
as the signature of absorption by ionized material along the line of sight
(Pounds et al. 1994, with \ros survey data; McHardy et al. 1995, 
with
\ros PSPC pointed data; Mihara et al. 1994, with \asca data; another study of \asca data 
by Guainazzi et al. is 
underway). The existence
of a warm absorber in NGC 4051 was firstly proposed by Fiore et al. (1992) as a possible explanation
for the observed spectral variability pattern in \ginga~data. 
Netzer et al. (1994)
found evidence for a warm absorber in NGC 4051 by introducing and applying X-ray 
colour diagrams.  

The overall spectral variability of NGC 4051 is complex with different behaviour
at different epochs, and no consistent description 
of the data has emerged yet. 
In general, there is a trend for the source to be softer when brighter (Lawrence et al. 1985,
Papadakis \& Lawrence 1995, with \exo~data; Matsuoka et al. 1990, with \ginga~data),  
traced back by Papadakis \& Lawrence (1995) to high flux peaks in the soft 
(0.05 - 2 keV) energy region only. 
Matsuoka et al. (1990) favour a correlation of powerlaw
index with the 2 - 10 keV luminosity, which Torricelli-Ciamponi \& Courvoisier
(1995) explain in terms of inverse Compton scattering of thermal UV photons by a population of nonthermal
relativistic electrons. 
Kunieda et al. (1992) and Fiore et al. (1992) describe part or the total 
of another \ginga~observation 
in terms of partial covering of the central source emitting a powerlaw with approximately constant slope. 

The optical spectrum of NGC 4051 
has been classified as either Seyfert 1.8 (e.g. Rosenblatt et al. 1992) or narrow-line Seyfert 1 (NLSy1 hereafter;
e.g. Malkan 1986). The line width (FWHM) of H$\beta$ is 990 km/s 
(Osterbrock \& Shuder 1982), broader than the forbidden lines with a
FWHM of [OIII]$\lambda$5007 of 330 km/s (De Robertis \& Osterbrock 1984). An additional weak broad 
component is seen in H$\beta$ (Veilleux 1991, Rosenblatt et al. 1994). 
High-ionization coronal lines are present in the optical and IR spectrum, like [FeXI]$\lambda$7892 
(De Robertis \& Osterbrock 1984) and [SiVI]1.96$\mu$m (Giannuzzo et al. 1995).  
   
The present X-ray observation 
was performed with the \ros PSPC
(Tr\"umper 1983; Pfeffermann et al. 1987) 
in order to study in detail the warm absorption feature, 
to discuss the time variability of the spectral components and to 
assess the possibility of a warm-absorber contribution to one of the emission-line components
seen in the IR to UV spectral region. 
 
The paper is organized as follows: In Sect. 2 we present the observations. In Sects. 3 and 4 the   
spectral and temporal analysis of the data is described and properties of the warm absorber
are derived. Sect. 5 is concerned with 
the discussion of the data; constraints on the unobserved EUV continuum are provided, the properties of 
the warm gas are discussed and further constrained, the absorber-intrinsic optical-UV emission
and absorption is investigated, the influence of dust is assessed, the thermal stability of the
warm material is examined, and the NLSy1 character of NGC 4051 is commented on.  
In Sect. 6 we provide a summary and the conclusions.  

A distance of 14 Mpc is adopted for NGC 4051, resulting from a Hubble constant of $H_{\rm o}$ = 50 km/s/Mpc
and the assumption that the galaxy follows the Hubble flow, i.e. any peculiar velocity component is small.
 
If not stated otherwise, cgs units are used throughout. 

  \begin{figure}[thbp]
      \vbox{\psfig{figure=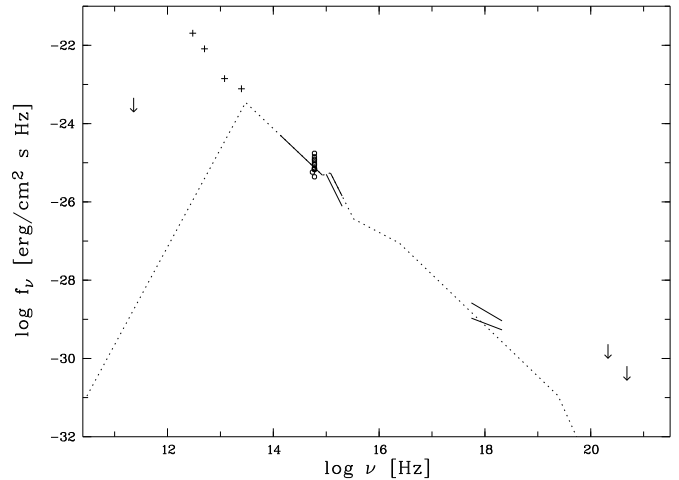,width=8.8cm,%
          bbllx=2.9cm,bblly=1.1cm,bburx=18.1cm,bbury=12.2cm,clip=}}\par
      \vspace{-0.4cm}
 \caption[SEDobs]{Extract of the observed spectral energy distribution of NGC 4051
 from the radio to the gamma-ray region, compiled from the literature (Edelson et al. 1987,
 Edelson \& Malkan 1986, Done et al. 1990, Malkan 1986, Rosenblatt et al. 1992, Walter et al. 1994,
 Maisack et al. 1995). Arrows
 denote upper limits. The circles correspond to measurements of the optical flux
 at 5000 \AA~at different epochs (Malkan 1986, Rosenblatt et al. 1992),
 giving an impression on the amplitude of the
 optical variability, which is similar in the UV as observed by IUE (Courvoisier \& Paltani 1992).
 The dotted line represents the spectrum chosen for modeling.
}
\label{SEDobs}
\end{figure}

\section{Data reduction}

The observation was performed with the \ros PSPC from November 11 -- 12, 1993, 
centered on NGC 4051.   
The total exposure time is about 12 ksec.   
The source photons were extracted 
within a circle chosen to be large enough to ensure that all of the source counts 
are included, given the detector response to very soft photons which causes
an extension of the image below 0.2 keV (Nousek \& Lesser 1993).  
The background was determined by removing all detected sources within the inner field of view.
The data were corrected for 
vignetting
and dead-time, using the EXSAS software (Zimmermann et al. 1994).
The mean source countrate is about 2.9 cts/s.
For the spectral analysis source photons in
the amplitude channels 11-240 were binned 
according to a constant signal/noise ratio of 16$\sigma$. 
For the temporal analysis the minimal binsize in time was 400 s to account for the
satellite's wobbling motion. 

\section{Spectral analysis} 

   \begin{table*}     
     \vspace{-0.5cm}
     \caption{X-ray spectral fits to NGC 4051 (pl = powerlaw, bb = black body,
                  wa = warm absorber). Model 3 refers to the one with
      `standard' assumptions (see text), in model 4 (and only in 4) the data corresponding to
        the short high-state in source flux (orbit 1) are excluded from the spectral fitting,
        in model 5 underabundant
       metals of 0.2 $\times$ solar are chosen and in model 6 an IR spectral component
       as observed by IRAS is added.
       The errors are quoted at the 95.5\% confidence level. 
        }
     \label{fitres}
      \begin{tabular}{llccllccc}
      \hline
      \noalign{\smallskip}
        model & log $N_{\rm H}$ & log $U$ & log $N_{\rm w}$ & log Norm$_{\rm pl}$
                            & $\Gamma_{\rm x}$ & log Norm$_{\rm bb}$ & $kT_{\rm bb}$
                            & $\chi^2(d.o.f)$ \\
       \noalign{\smallskip}
      \noalign{\smallskip}
           & [cm$^{-2}$] & & [cm$^{-2}$] & [ph/cm$^2$/s/keV] & & [ph/cm$^2$/s] & [keV] & \\
       \noalign{\smallskip}
      \hline
      \hline
      \noalign{\smallskip}
 1~~~pl & 20.24$^{+0.04}_{-0.04}$ & - & - & --2.44$^{+0.01}_{-0.01}$$^{(1)}$
                & --2.92$^{+0.07}_{-0.07}$ & - & - & 230(59) \\
      \noalign{\smallskip}
      \hline
      \noalign{\smallskip}
 2~~~pl + bb & 20.11$^{(3)}$ & - & - & --2.51$^{+0.01}_{-0.01}$ & --2.6$^{+0.1}_{-0.1}$
                      & --1.67$^{+0.09}_{-0.13}$ & 0.10$^{+0.01}_{-0.01}$ & 110(58) \\
      \noalign{\smallskip}
      \hline
      \noalign{\smallskip}
  3~~~wa & 20.11 & 0.40$^{+0.03}_{-0.03}$ & 22.67$^{+0.02}_{-0.02}$
               & --4.49$^{+0.01}_{-0.01}$$^{(2)}$ & --2.3$^{(4)}$ & - & - & 67(59) \\
      \noalign{\smallskip}
      \hline
      \noalign{\smallskip}
  4~~~wa, no orb1 & 20.11 & 0.40$^{+0.03}_{-0.03}$ & 22.67$^{+0.02}_{-0.02}$
               & --4.52$^{+0.01}_{-0.01}$ & --2.3 & - & - & 62(59) \\
      \noalign{\smallskip}
      \hline
      \noalign{\smallskip}
  5~~~wa, met=0.2 & 20.11 & 0.60$^{+0.03}_{-0.03}$ & 23.38$^{+0.02}_{-0.02}$
               & --4.49$^{+0.01}_{-0.01}$ & --2.3 & - & - & 70(59) \\
      \noalign{\smallskip}
      \hline
      \noalign{\smallskip}
  6~~~wa, IRAS-IR & 20.11 & 0.20$^{+0.03}_{-0.03}$ & 22.69$^{+0.02}_{-0.02}$
               & --4.50$^{+0.01}_{-0.01}$ & --2.3 & - & - & 70(59) \\
      \noalign{\smallskip}
      \hline
      \noalign{\smallskip}
  \end{tabular}

\noindent{\small $^{(1)}$ Normalization at 1 keV ~~~ $^{(2)}$ at 10 keV ~~~ $^{(3)}$ fixed to the Galactic value
~~~ $^{(4)}$ pre-determined and then fixed (see text)
}
   \end{table*}
\subsection{Simple spectral models}
A single powerlaw has been previously found to give a poor fit to the X-ray spectrum of
NGC 4051.
This also holds for the present observation (see Table \ref{fitres}). The resulting 
powerlaw slope is rather steep, $\Gamma_{\rm x} \approx -2.9$, and 
strong residuals remain, with an unacceptable $\chi{^{2}}_{\rm red}$ of 3.8.  
When the cold absorbing column density $N_{\rm H}$ is left as a free parameter, a value slightly  
higher than the Galactic one is found.  
For all further fits described below, $N_{\rm H}$ turned out to be either
less than or about the same as the Galactic value and thus was fixed to the Galactic
column of 0.13 $ \times 10^{21}$ cm$^{-2}$ (Elvis et al. 1989; see also 
McHardy et al. 1995).   

Although an improvement of the fit is obtained when adding a soft excess, this
model does not provide a satisfactory description of the spectrum, either. 
Parameterizing the excess component
as a black body (with normalization and temperature free) yields  
$\chi^{2}_{\rm red}$ = 1.9 with a rather steep underlying powerlaw spectrum (Table \ref{fitres}). 

Alternatively, the fit can be improved by adding an absorption edge
which is thought to originate from warm gas along the line of sight.  
With the resulting best fit from a physical warm absorber model (next section) 
already in mind, the major absorption edges being those of OVII, OVIII and NeX,
we modeled the X-ray spectrum with 3 single edges (of the form $\phi(E,\tau)=e^{-\tau(E/E_{\rm edge})^{-3}}$
for $E \ge E_{\rm edge}$) 
superimposed on a powerlaw. The edge energies were fixed to the theoretical values 
($E_{\rm OVII}$ = 0.74 keV, $E_{\rm OVIII}$ = 0.87 keV, $E_{\rm NeX}$ = 1.36 keV) 
and the optical depths $\tau$ left free to vary. We find $\tau_{\rm OVII} = 0.35\pm0.15$, 
$\tau_{\rm OVIII} = 1.1\pm0.4$, $\tau_{\rm NeX} = 0.8\pm0.4$ and a photon index $\Gamma_{\rm x} = -2.40\pm0.06$.    

\subsection{Warm absorber models}  

The spectral absorption structure resulting from 
a physical absorber is more complex than a simple edge.  
Assuming the gas to be photoionized by continuum emission of the central
pointlike nucleus and assuming it to be in photoionization equilibrium, 
we calculated a sequence of photoionization models
using the code {\em Cloudy} (Ferland 1993). 

The ionization state of the warm absorber can be characterized by the
hydrogen column density $N_{\rm w}$ of the warm material and the
ionization parameter $U$, defined as  
\begin {equation}
U=Q/(4\pi{r}^{2}n_{\rm H}c)
\end {equation} 
where $Q$ is the
number rate of incident photons above the Lyman limit, $r$ is the distance between
central source and warm absorber, $c$ is the speed of light, and 
$n_{\rm H}$ is the hydrogen density 
(fixed to 10$^{9.5}$ cm$^{-3}$ unless noted otherwise).  
The X-ray absorption structure depends only very weakly on $n_{\rm H}$ ($\approx$ $n_{\rm e}$,
where $n_{\rm e}$ is the electron density) and 
remains unchanged for the range of densities discussed below, given the dominant physical
processes; the essential parameter is the 
column density $N_{\rm w}$. 
Both quantities, $N_{\rm w}$ and $U$, are determined from the X-ray
spectral fits. 
Solar abundances (Grevesse \& Anders 1989)
were adopted if not stated otherwise.
We have always assumed the continuum source to be completely covered by the absorber,
i.e. no partial covering in which part of the intrinsic X-ray spectrum is seen directly, 
has been studied. Further, we assume the absorber to be one-component. 
 
The observed spectral energy distribution (SED) of NGC 4051
is shown in Fig.  \ref{SEDobs}. 
The one chosen for the modeling corresponds to the 
observed IR to UV spectrum in an intermediate brightness state
(corrected for stellar contribution) 
extrapolated to the Lyman limit, 
a break at
10$\mu$m and an energy index $\alpha$ = --2.5 $\lambda$-longwards, an X-ray powerlaw
as self-consistently determined from the spectral fitting, and a break into the
gamma-ray region at 100 keV.  

Data in the NIR to UV spectral region were taken from Done et al. (1990) and are
simultaneous 
but not contemporaneous to the present X-ray observation, although both represent
an intermediate brightness state.  Anyway, Done et al. (1990) and 
Hunt et al. (1992) find no short-timescale correlated 
IR/optical/UV -- X-ray variability. Long-term trends are less certain,
but Salvati et al. (1993) point to an X-ray high-state 
contemporaneous with a NIR high-state.~~~ 
The FIR spectrum of NGC 4051 observed by IRAS does not show a 
turnover shortwards of 100 $\mu$m. 
However, the IRAS points are thought to be 
contaminated 
by emission from cold dust of the surrounding galaxy (Edelson \& Malkan 1986, Ward et al. 1987) 
and consequently are not seen by the warm material as a pointlike continuum  
source. Therefore, the nuclear SED was assumed to break at 10 $\mu$m and
extend to the radio spectral region with $\alpha$ = --2.5,   
consistent with the measured millimeter upper limit (Edelson et al. 1987). 
Anyway, the continuum $\lambda$-longwards the Lyman limit usually does not play an important
role in determining the ionization state of the warm material and in particular the
depth of the X-ray absorption features. (Later    
we comment on the influence of increased  
free-free heating, when an IR component as observed by IRAS is added to the nuclear 
spectrum.)    
The hard X-ray spectrum was assumed to follow the soft X-ray powerlaw 
(as was found to be a viable description of the \ros--\ginga and \asca data; 
Pounds et al. 1994, Mihara et al. 1994, Guainazzi et al. 1996)
and to break
at 100 keV, in line with current observations of Seyferts (e.g. Kurfess 1994) and consistent
with the observed gamma-ray upper limit (Maisack et al. 1995). 

\begin{figure}[thbp]
      \vbox{\psfig{figure=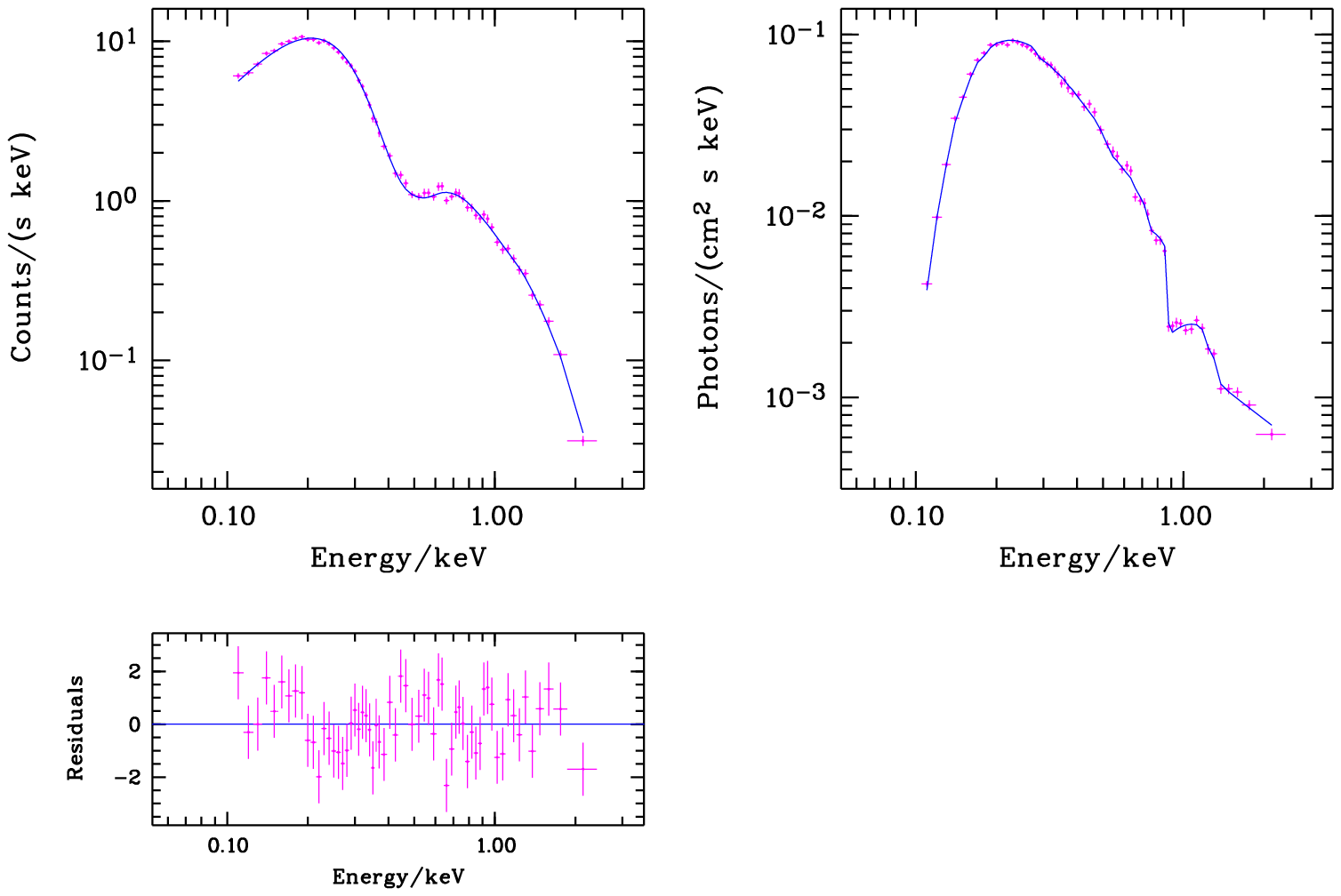,width=8.8cm,%
          bbllx=2.5cm,bblly=1.1cm,bburx=10.3cm,bbury=4.5cm,clip=}}\par
      \vspace{-0.4cm}
\caption[SEDx]{Residuals of the warm absorber fit to the X-ray spectrum of NGC 4051.
The fit parameters
are listed in Table \ref{fitres} and the unfolded spectrum is shown in Fig. \ref{SEDfin}.}
\label{SEDx}
\end{figure}

Fitting the warm absorber model to the X-ray spectrum of 
NGC 4051 results in an ionization parameter of $\log U = 0.40 \pm 0.03$, a warm 
column density of $\log N_{\rm w} = 22.67 \pm 0.02$, and a photon index of 
$\Gamma_{\rm x}$ = --2.3 (Table \ref{fitres}). The value of $\Gamma_{\rm x}$ 
was pre-determined in a sequence of fits and then fixed to that value
for further modelling (we did not attempt to tune the second digit of 
$\Gamma_{\rm x}$; a change of $\Delta\Gamma_{\rm x} = +0.1 (-0.1)$ 
results in $\Delta\chi^{2}$ = 24(10)).  
Further discussions and conclusions refer to this set of model parameters, if
not stated otherwise.
The electron temperature $T$ of the warm gas is about 3 $\times$ 10$^5$ K.  
There is no evidence for a cold absorbing column larger than the Galactic one.  
The residuals from the best spectral fit are shown in Fig. \ref{SEDx} and the
(unfolded) X-ray spectrum is displayed in Fig. \ref{SEDfin}. 
The absorption structure is dominated by highly ionized oxygen (OVIII) and neon.  
No strong iron edge around 7 keV
is predicted (Fig. \ref{SEDfin}), consistent with higher-energy observations (e.g. Matsuoka et al. 1990). 
The intrinsic X-ray luminosity (i.e. the one prior to warm absorption) for this model
in the 0.1 -- 2.4 keV energy range
is $L_{\rm x} = 9.5 \times 10^{41}$ erg/s. 
%

Netzer (1993) has discussed the consequences for the X-ray spectral shape 
of taking into account X-ray emission 
and reflection by the ionized absorbing gas.  
For the present observation of NGC 4051 the addition of 
an emission and reflection component to the X-ray spectrum, calculated  
with the code {\em Cloudy} for a covering factor of the warm material of 0.5, only negligibly changes  
the results ($\log N_{\rm w}$ = 22.70) due to the weakness of these components.

The warm absorber fit shown in Fig. \ref{SEDx} shows some residual structure 
between 0.1 and 0.3 keV. For completeness we note that an additional very soft
excess component, parameterized as a black body, removes part of the residuals.
Of course, 
the parameters of such a black body component are not well constrained from X-ray spectral
fits. One with $kT_{\rm bb}$ = 13 eV (corresponding to $T_{\rm bb} \approx$ 150\,000 K) 
and an integrated absorption-corrected flux (between the Lyman limit and 2.4 keV)
of $F = 7 \times 10^{-11}$ erg/cm$^2$/s fits the data. It contributes about the same 
amount to the ionizing luminosity as the powerlaw continuum used for the modeling 
and has the properties of the EUV spectral component for which evidence is presented in Sect. 5.1.  

The observational data
were compared to  
four further model sequences, which consisted of 
(i) a change in metal abundances of the warm gas up to 0.2 $\times$ solar, 
(ii) the addition of dust with Galactic (ISM) properties to the gas, or a modification
of the SED impinging on the absorber, by 
(iii) the addition of an EUV black body component with a temperature of 
$T_{\rm bb}$ = 100\,000 K or 150\,000 K contributing
the same amount to the ionizing luminosity as the powerlaw component  
(the choice of these parameters is motivated
and further discussed in Sects. 5.2.1, 5.2.4 and 5.1, respectively), or (iv) the inclusion of a strong IR spectral
component as observed by IRAS. 

The resulting best-fit model for reduced metal abundances of 1/5 the 
solar value is given in row 4 of Table 1. The low abundances are mostly reflected in
an increase in the corresponding total hydrogen column $N_{\rm w}$, as expected due to the
fact that oxygen (and neon) are most important in determining the absorption structure. 
An additional IR component strongly increases the free-free heating of the
gas and the electron temperature rises. 
An additional
black body component in the EUV with the properties given above has negligible
influence on the X-ray absorption structure.
Dusty models will be further commented on in Sect. 5.2.4.  

\subsection {Warm absorber plus soft excess ?}

The intrinsic, i.e. unabsorbed, powerlaw with index $\Gamma_{\rm x}$ = --2.3 is steeper
than the Seyfert-1 typical one with --1.9.
In fact, fixing the slope of the intrinsic powerlaw to $\Gamma_{\rm x}$ = --1.9
(with $U$ and $N_{\rm w}$ as free parameters) leads to a significantly
worse fit ($\chi{^{2}}_{red}$ = 5.2). This still holds when 
all data points below 0.3 keV (which show some residual structure
interpretable as a very soft excess, as mentioned above) 
are excluded from the spectral fitting ($\chi{^{2}}_{red}$ = 4.3).   

Since a soft excess on top of a flat powerlaw might mimick a single steeper
powerlaw, we have performed some further tests to check, whether the data
can be reconciled with $\Gamma_{\rm x}$ = --1.9 plus a soft excess, parameterized
as a black body.  
Firstly, we have fixed the black body temperature to the value found in an \asca
observation by 
Mihara et al. (1994), $kT_{\rm bb}$ = 0.1 keV. In this case, the black
body contribution is always found to be negligible (with a normalization 
of less than 10$^{-10}$ ph/cm$^2$/s). 
When the other fit parameters, namely the ionization parameter of the warm absorber, 
are changed to enforce a black body contribution to the fit, $\chi^{2}$ remains far
above acceptable values. 
Secondly, $T_{\rm bb}$ was left as an additional free parameter. In this case, a
very soft excess is found, with $kT_{\rm bb} \approx$ 35 eV, ill-constrained 
by the \ros data, and the overall quality of the fit is not yet acceptable. 
Systematically steepening the underlying powerlaw, one is lead back to 
the model presented above, with $\Gamma_{\rm x}$ = --2.3. 
Finally, all data points which correspond to the short {\em high-state} in source flux
and which show some evidence for an additional soft excess ($kT_{\rm bb} \approx$ 0.12 keV; detailed
in Sect. 4.2) were excluded from the spectral fitting. The underlying powerlaw
spectrum remains steep and the deduced parameters of the warm absorber remain
unchanged within the error bars (Table 1, model 4). 

\section {Temporal analysis}

\subsection {Flux variability}

\begin{figure}[thbp]
      \vbox{\psfig{figure=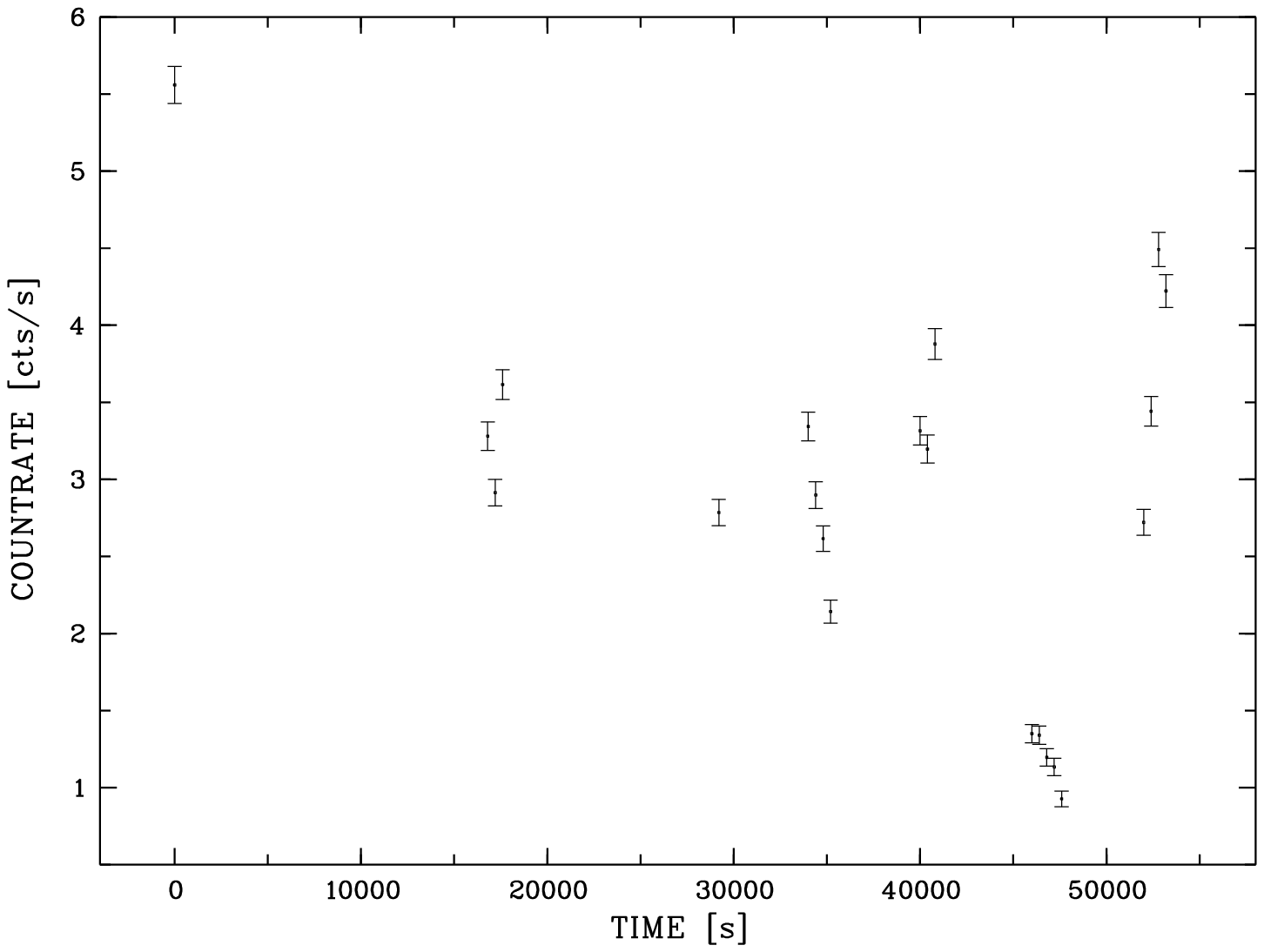,width=8.8cm,%
          bbllx=3.2cm,bblly=1.1cm,bburx=18.1cm,bbury=12.2cm,clip=}}\par
      \vspace{0.4cm} 
      \vbox{\psfig{figure=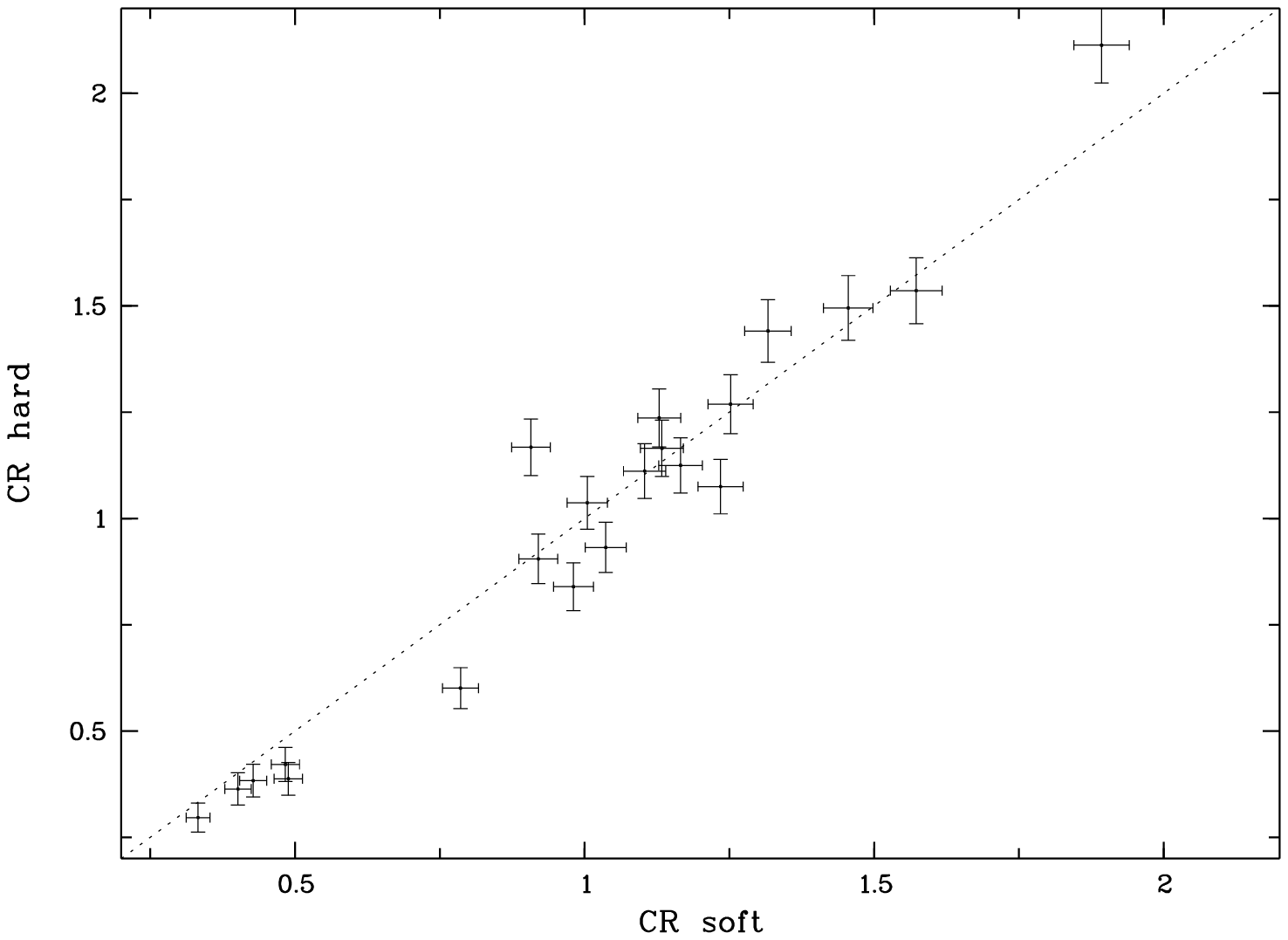,width=8.8cm,%
          bbllx=3.1cm,bblly=1.1cm,bburx=18.1cm,bbury=12.2cm,clip=}}\par
      \vspace{-0.4cm}
 \caption[light]{(a) X-ray lightcurve of NGC 4051, binned to time intervals of 400 s for the total
energy range.\\ 
(b) Correlation of soft (0.1-0.5 keV) and hard (0.5-2.4 keV) countrate normalized to
the mean countrate in the corresponding energy interval. The dotted line represents  
a linear relation. }
 \label{light} 
\end{figure}

NGC 4051 is known to be rapidly variable.  
The X-ray light curve
of the present observation is shown in Fig. \ref{light}.
%
The source is variable
by about a factor of 6 within a day.  
In order to test whether the amplitude of variability is the same in the low and high energy 
region of the \ros band, we have divided the total observed flux in a soft (0.1 keV $\le E \le$ 0.5 keV) and a hard
(0.5 keV $\le E \le$ 2.4 keV) component and normalized each to the mean flux in the corresponding
band.
(We note that only in this section do we use this `tight' definition of `soft' band. 
Throughout the rest of the paper, the term `soft excess' more losely refers to a component
somewhere in the \ros energy band without implying it to be located exclusively below 0.5 keV.)  
The soft band is dominated by the cold-absorbed powerlaw
component, whereas the dominant 
feature of the warm absorber, the oxygen absorption edge,
is located in the hard band. 
We find essentially correlated variability
between both components (Fig. \ref{light}b), although there is a slight trend for the
source to be softer when fainter. 
   
\subsection{Spectral variability}

\begin{figure}
      \vbox{\psfig{figure=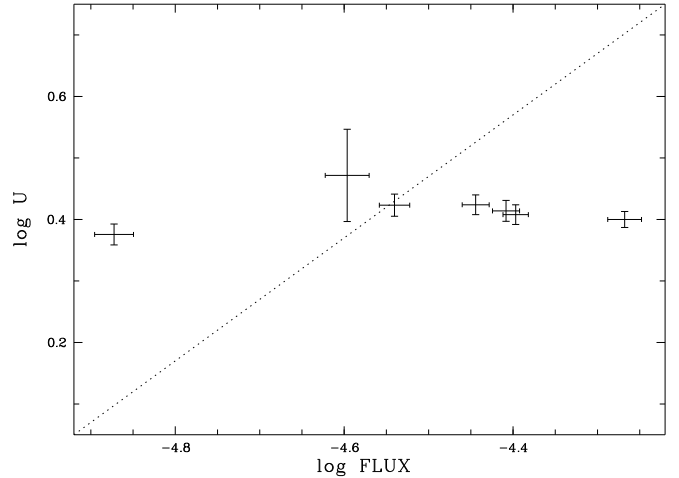,width=8.8cm,%
          bbllx=3.0cm,bblly=1.1cm,bburx=18.1cm,bbury=12.2cm,clip=}}\par
      \vspace{-0.4cm}
 \caption[Uvar]{Ionization parameter $U$ in dependence of the intrinsic X-ray
luminosity (parameterized as the powerlaw flux at 10 keV, in photons/cm$^2$/s/keV),
resulting from warm absorber
fits to individual orbits of the total observation. The errors correspond to
95.5 \% confidence.
The dashed line reflects the expected dependence of $U$ on flux if
the warm gas reacted immediately to changes in the ionizing luminosity.
Within the error bars, $U$ is constant throughout the total observation.}
 \label{Uvar}
\end{figure}

\begin{figure}[thbp]
      \vbox{\psfig{figure=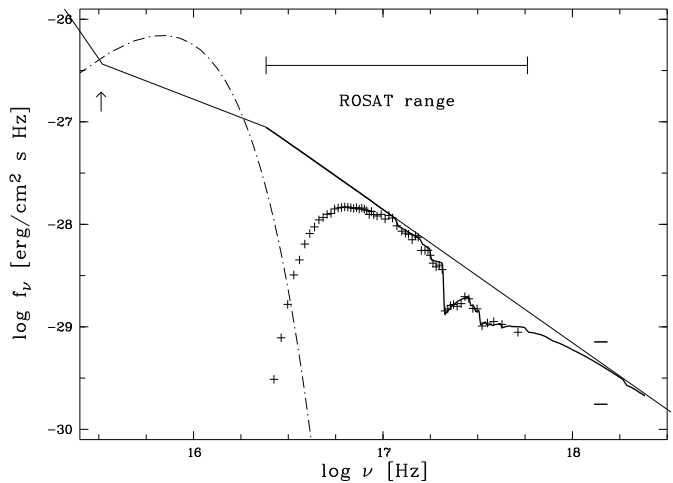,width=8.8cm,%
          bbllx=2.9cm,bblly=1.1cm,bburx=18.1cm,bbury=12.2cm,clip=}}\par
      \vspace{-0.4cm}
\caption[SEDfin]{UV- to X-ray spectrum of NGC 4051. The thin solid line corresponds
to the unabsorbed SED chosen for modeling. The thick solid line in the X-ray region shows the
best-fit warm-absorbed spectrum corrected for Galactic cold absorption, whereas the crosses
correspond to the absorbed spectrum. 
The dot-dashed line represents the black body with $kT$ = 13 eV
that was mentioned in Sect. 3.2.  
The arrow marks the Lyman limit and the two horizontal bars indicate the amplitude of
flux variability when the data are binned to orbits.}
\label{SEDfin}
\end{figure}

To check for variability of the
warm absorption feature in more detail, we have performed warm-absorber fits to individual subsets of the
total observation (referred to as `orbits' hereafter). The warm column density, $N_{\rm w}$, 
is not expected to vary on short timescales
and was fixed to the value determined for the
total observation. The same was done for the unabsorbed powerlaw index $\Gamma_{\rm{x}}$ = --2.3
(but see comment below).
The best-fit ionization parameter turns out to be essentially constant
over the whole observation despite strong changes in the intrinsic luminosity (Fig. \ref{Uvar}).
If the warm material reacted instantaneously to variations in the ionizing luminosity,
a clear correlation between $U$ and $L$ would be expected. Given the fact that there are
time gaps in the observation and slightly delayed reactions might have escaped observation, one
would still expect $U$ to scatter as strongly as $L$.    
The constancy of $U$ provides a limit on the density of the warm gas. 
Its recombination timescale $t_{\rm{rec}}$
is conservatively estimated from the lack of any reaction of the warm material
during the long low-state in orbit 7 (at $t$ = 46\,000 -- 48\,000 s after $t_{\rm start}$; cf. Fig. 3a), 
resulting in $t_{\rm{rec}} \approxgt 2000$ sec. 
The upper limit on the density is given by 
\begin{equation}
n_{\rm{e}} \approx {1\over t_{\rm{rec}}}~{n_{\rm{i}}\over n_{\rm{i+1}}}~{1\over A}({T\over 10^4})^{X}
\end{equation} 
where $n_{\rm{i}}/n_{\rm{i+1}}$ is the ion abundance ratio of the metal ions 
dominating the cooling of the gas 
and the last term is the 
corresponding recombination rate coefficient $\alpha_{\rm{i+1,i}}^{-1}$ (Shull \& Van Steenberg 1982
with coefficients $A$ = 6.71 $\times 10^{-11}$, $X$ = 0.726 from Aldrovandi \& Pequignot 1973).
Oxygen is  
a major coolant, and given its ionization structure for the best-fit
warm absorber model  
(with an ion abundance ratio of $n_{\rm{O^{7+}}}/n_{\rm{O^{8+}}} \approx 0.3$ and 
a recombination rate coefficient of $\alpha$ = 0.57 $\times 10^{-11}$ cm$^{3}$/s) 
this yields $n_{\rm e} \approxlt$ $3 \times 10^{7}$cm$^{-3}$. 

For the above analysis, the intrinsic powerlaw index was fixed to
the value derived for the total observation, $\Gamma_{\rm{x}}$ = --2.3. In fact, when fitting
the X-ray spectra of individual orbits, this value always turns out to still represent the
best description of the data.  
Due to the lower number
of photons in each orbit, these datasets are less restrictive
concerning the values of the fit parameters.
However, Fig. 3b already indicates that no strong spectral changes 
occured, although simultaneous variations of different components that 
simulate a constant countrate ratio cannot be excluded.   
To set limits on the variability of $\Gamma_{\rm x}$ within the present observation 
we have fixed the powerlaw index to --1.9. 
Re-fitting the low-state (a hardening of the spectrum when the
source is less bright would mirror the trend that is observed at higher energies,
e.g. Matsuoka et al. 1990), no satisfactory description of the data can be
achieved, with a change $\Delta\chi^2$ = 39 as compared to the former best fit. 
  
The spectrum during the first orbit (i.e. the first $\approx$ 800 s) of the total observation,
corresponding to a high-state in source flux,
shows evidence for an additional black-body-like spectral component
with $kT_{\rm bb} \approx$ 0.12 keV and an unabsorbed flux of 7.7 $\times~10^{-12}$ erg/cm$^2$/s in the \ros
band, consistent with former observations of a soft excess during 
source high-states (Pounds et al. 1994, Mihara et al. 1994).    
If, again, $\Gamma_{\rm{x}}$ = -- 1.9 is enforced, and the parameters of
the black body are left free to vary, an additional very soft excess
with $kT_{\rm bb} \approx$ 40 eV reproduces the observation. 
However, a very large column of the warm material is found in this case
to compensate the flatter intrinsic powerlaw, log $N_{\rm w}$ = 23.4.  
If this model applied, a strong change in the column density (by about a factor of 5) during the high-state
would have to be invoked, which seems rather unrealistic.   
%

\section{Discussion}

\subsection{SED and ionization parameter}  

The ionization parameter of the warm absorber, as determined from 
the X-ray absorption spectrum, is $\log U = 0.4$.
Here, the EUV-SED incident on the absorber 
enters in two ways: (i) by contributing to the total number
of photons irradiating the warm gas (Eq. (1)) and (ii) by its influence on the
ionization structure of the absorber via its spectral shape. 
In the following, we discuss constraints on the EUV-SED and the
resulting number rate $Q$ of photons above the
Lyman limit.
It should be noted, however, that the X-ray part of the SED is the one most 
important in determining the depth of the metal absorption edges, in creating 
highly ionized ions by K-shell photoionization. 

A lower limit 
on the number of photons in the unobserved EUV-part of the SED is estimated 
by a powerlaw interpolation between the
flux at 0.1 keV (from X-ray spectral fits) and the Lyman-limit (by extrapolating
the observed UV spectrum), which was used for the present modeling. 
This gives $Q = 1.6 \times 10^{52}$ s$^{-1}$.

$Q$ can also be deduced from the H$\beta$ luminosity. The minimal number of hydrogen-ionizing photons
isotropically emitted by the central continuum source is given by the total
observed H$\beta$ luminosity. Assuming $T=(10-20) \times 10^3$ K and
using Tables. 2.1 and 4.2 of Osterbrock (1989)
results in $Q = (2.1-3.8)\times$10$^{12}$ $L_{\rm H\beta}$. The mean observed $L_{\rm H\beta}$ =
7.8 $\times 10^{39}$ erg/s (Rosenblatt et al. 1992) yields
$Q = (1.6-2.5) \times 10^{52}$ s$^{-1}$. It is interesting to note that this is of the same order
as the lower limit determined from the assumption of a single powerlaw EUV-SED, suggesting
the existence of an additional EUV component in NGC 4051. (Alternatively, it would imply
that broad line region (BLR) plus narrow line region (NLR) completely cover the central
source (i.e. the covering factor is unity) 
in contrast to what is observed in the soft X-ray spectrum, i.e. there is 
no evidence for strong cold absorption in excess of the Galactic value.)  
However, such an EUV component cannot be identified with the black-body-like soft X-ray excess
seen in source high-states (Sect. 4.2; Pounds et al. 1994, Mihara et al. 1994) which
turns over already in the soft X-ray region, negligibly contributing to the EUV
luminosity. 
Neither is there a measurable black-body continuum component in the ultraviolet 
(Edelson \& Malkan 1986). 

A third approach to $Q$ is via the ionization-parameter-sensitive
emission-line ratio [OII]$\lambda$3727/[OIII]$\lambda$5007 (e.g. Penston et al. 1990).
This method makes use of the fact that the [OII]/[OIII] ratio is rather insensitive
to the spectral shape of the ionizing continuum 
(although it tends to overestimate the number of photons, if the emission line
region in question contains density inhomogeneities or a large fraction of matter-bounded
clouds; Schulz \&~Komossa~ 1993).  The oxygen ratio, as measured by Malkan (1986)  
yields an ionization parameter for the NLR of $\log~U_{\rm{NLR}} = -2.2$ - 
--2.5. 
To determine $Q$, the mean density and distance of the narrow line region have to be known. 
The density was estimated from the emission line ratio [SII]$\lambda$6716/[SII]$\lambda$6731
(Veilleux 1991) to be $n_{\rm H} = 6 \times 10^{2}$ cm$^{-3}$ and the distance 
assumed to be $r = 50$ pc (consistent with Schmitt \& Kinney 1996), 
leading to $Q = (1.7 - 3.4) \times 10^{52}$ s$^{-1}$.

An additional EUV component may also explain 
the observational trend that broad emission lines and observed continuum 
seem to vary independently in NGC 4051 
(Osterbrock \& Shuder 1982, Peterson et al. 1985, Rosenblatt et al. 1992), 
as expected if the observed optical-UV continuum
variability is not fully representative of the EUV regime.

As noted in Sect. 3.2 there are hints for such an EUV bump in the
present X-ray spectrum, although the evidence is rather weak due to the softness of the component. 
A similar very soft excess, based on much better photon statistics, has been found in the \ros spectrum of 
the narrow-line quasar TON\,S180 (Fink et al. 1996).   

Given the evidence for an EUV bump in NGC 4051 in excess of a simple
powerlaw,
we have verified
that such a component (parameterized as a black body; Sect. 3.2) does not change
the fit parameters of the warm absorber within the given error bars except for 
contributing to $U$. 

An intense IR spectral component, however, leads to a strong heating of the
gas and the best-fit ionization parameter decreases somewhat.
Underabundant metals (see next section) lead to an increase in the fit value of $U$.
There is approximately a factor of 2 uncertainty in $U$ resulting
from this ignorance of continuum shape and chemical composition.

\subsection{Properties of the warm absorber} 

\subsubsection {Column density and abundances}
The warm hydrogen column density of the absorber is $\log N_{\rm w}$ = 22.7, 
adopting solar abundances; an assumption that is usually made.
The actual abundances are a priori unknown and might
deviate from that by a factor of several. 
There are indications for underabundant metals in the extended emission line regions of
some Seyfert galaxies (e.g. Tadhunter et al. 1989) and in the narrow line regions 
of Seyfert 2 galaxies by a factor of 1/2 to 1/3
compared to the solar value (Komossa \& Schulz 1994).  
Marshall et al. (1993) constructed a model for the warm electron scattering medium
in NGC 1068 (which, in general, might be one component to identify the warm absorber with; and
the temperature of which is comparable to that found for the absorber in NGC 4051) 
with underabundant oxygen of 1/5 $\times$ solar. 
Depleted gas-phase oxygen abundances are also found
by Sternberg et al. (1994) in molecular gas in the central region of NGC 1068, 
and material evaporating from the torus  
might be another reservoir for warm absorbing material. 
X-ray spectral fits do not constrain the abundances.    
An approach to estimate the chemical composition 
within the inner region of active galaxies 
is by emission lines.  
For the narrow line region, the intensity 
ratio [OIII]$\lambda$4363/[OIII]$\lambda$5007 is a good 
abundance indicator via its temperature sensitivity. Using published emission line intensities
of NGC 4051 (Dibai \& Pronik 1968, Malkan 1986),
we find $\log$ [OIII]$\lambda$4363/[OIII]$\lambda$5007 $\approx$ --0.7, 
indicative of a rather high temperature and correspondingly low abundances. 
 
Running fits with reduced metal abundances of 1/5 $\times$ solar results
in a larger total warm column density, $\log N_{\rm w}$ = 23.38, reflecting the fact 
that the depth of the absorption feature is dominated by oxygen.

\subsubsection {Density}  
The density of the warm gas is not important in determining the X-ray spectral shape
at a fixed time. 
A limit was drawn from
the variability behaviour (Sect. 4.2), which resulted in $n_{\rm{H}} \approxlt$ $3 \times 10^{7}$cm$^{-3}$.
We come back to this point in 
Sect. 5.3.1.  

With $n_{\rm{H}} \le$ $3 \times 10^{7}$cm$^{-3}$, the thickness of the warm absorber is
$D \ge 2 \times 10^{15}$ cm. 
 
\subsubsection {Location}     

The location of the warm material is poorly constrained from X-ray spectral
fits alone. The absorber might be part of the   
broad line region or situated
farther outwards, e.g. in the narrow line region.
The emission line contribution of an optically thin
matter-bounded BLR component in active galaxies was discussed by Shields et al. (1995),
who also particularly pointed out that
this gas might act as an X-ray warm absorber.

Using $Q = 1.6 \times 10^{52}$ s$^{-1}$ as derived from 
the powerlaw description of the EUV spectrum
and the upper limit of the density of the warm
gas as determined from the X-ray variability results in a distance of the absorber
from the central power source of $r \ge 3 \times 10^{16}$ cm. 
Conclusive results for the distance of the BLR in NGC 4051 from
reverberation mapping, 
that would allow a judgement of the relative positions of both components, 
do not yet exist: Rosenblatt et al. (1992) find the optical continuum 
to be variable with large amplitude, but no significant change in the H$\beta$ flux.  

\subsubsection {Influence of dust} 
Dust might be expected to survive in the warm absorber, e.g. depending on its distance
from the central energy source. 
A rough estimate for the evaporation distance $r_{\rm ev}$
of dust is provided by $r_{\rm ev} \approx \sqrt{L_{46}}$ pc, where $L$ is the integrated
continuum luminosity in 10$^{46}$ erg/s (Netzer 1990),
leading to $r_{\rm ev} \approx 0.02$ pc for NGC 4051. 

Mixing dust of Galactic ISM properties (including both, graphite and  
astronomical silicate; Ferland 1993) with the warm gas in NGC 4051 
and self-consistently re-calculating the models 
leads to maximum dust temperatures of 2200\,K (graphite) and 3100\,K (silicate), above the evaporation 
temperatures (for a density of $n_{\rm H} = 5 \times 10^7$ cm$^{-3}$ and $U$, $N_{\rm w}$ 
of the former best-fit model; Sect. 3.2).  
For $n_{\rm H} \approxlt 10^6$ cm$^{-3}$, dust can survive throughout the absorber.
However, it strongly changes the equilibrium conditions and 
ionization structure of the gas via strong photoelectric heating and collisional cooling. 
For relatively high ionization parameters,
dust very effectively competes with the gas
in the absorption of photons (e.g. Laor \& Draine 1993). 

An interesting point to pursue in this context is the following: An old puzzle is the
lack of any transition region between BLR and NLR (judged from missing `intermediate'
line emission). Netzer \& Laor (1993) proposed this to be due to the relatively more
important influence of
dust in an intermediate zone. Can the warm absorber be identified with this transition region ? 

Firstly, re-running a large number of models, we find no successful fit of the X-ray spectrum. 
This can be traced back to the relatively higher importance of edges from more lowly
ionized species, significantly changing the X-ray absorption spectrum 
and particularly a very strong Carbon edge.
Secondly, independent evidence for non-dusty warm gas comes from the observed UV and EUV spectrum,
if these components travel along the same path as the X-ray spectrum. The 2175 \AA ~bump 
in the UV spectrum of NGC 4051
is not particularly strong (e.g. Walter et al. 1994) and NGC 4051 is detected by EUVE (Marshall et al. 1995).    
(There are various possibilities to change the properties of dust mixed with the warm material.
The one which minimizes the aforementioned observable features, i.e. weakens the 2175 \AA ~absorption
and the 10$\mu$ IR silicon feature, and is UV gray, consists of   
a modified grain size distribution, with 
a dominance of larger grains (Laor \& Draine 1993). However, again, such models 
do not fit the observed X-ray spectrum, even if silicate only is assumed to avoid a strong
carbon feature and its abundance is depleted by 1/10.)     
Indubitably, the dust in NGC 4051 could be significantly different
from the Galactic one.
However, too many additional free parameters are introduced in this case to warrant a more
detailed study.
 
We conclude that the soft X-ray spectrum of NGC 4051 is not 
dominated by a (Galactic-ISM-like){\em dusty} warm absorber. We emphasize, however, 
that dusty warm absorbers might explain the low-energy absorption edges 
seen in some active galaxies (work in progress, for first results see Komossa \& Fink 1996). 
   
The absence of dust in the warm material would imply  
either (i)  the {\em history} of the warm gas is such that dust was never able to
              form,  like e.g. in an inner-disc driven outflow (e.g. K\"onigl \& Kartje 1994, Murray et al. 1995,
               Witt et al. subm.)
    or     (ii) if dust originally existed in the absorber,  the {\em conditions in} 
                the gas have to be such that dust destruction 
                 is guaranteed. In the latter 
                 case,  one obtains an important constraint on the density (location) of
                 the warm gas, which then has to be high enough (near enough) to ensure 
                 the dust is destroyed. For the present case (and Galactic-ISM-like dust) 
                 only a narrow range in density around
                 $n_{\rm H} \approx 5 \times 10^7$ cm$^{-3}$ is allowed. For lower densities, 
                 dust can survive in at least part of the absorber and higher densities have already 
                 been excluded in Sect. 4.2.     

\subsubsection {Warm-absorber intrinsic line emission and covering factor}

Constraints on the covering factor of the warm gas result from (i) its 
emissivity in emission lines, which has to be low enough not to predict 
line emission 
stronger than the observed one, and (ii) the desire to account for 
the rather large number of Seyferts which show evidence for warm absorption
(e.g. Fabian 1996),
leading to the expectation of a correspondingly large {\em mean} covering. 

In particular, it is interesting to ask whether one of the emission line
regions present in NGC 4051 (and in Seyfert galaxies in general), 
like the coronal line region, or the one responsible 
for the broad component seen in H$\beta$ in NGC 4051,
can be identified with the warm absorber.  
 
In the case of NGC 4051, the maximal warm-absorber intrinsic H$\beta$ 
emission predicted for the best-fit model is
only about 1/220 of the observed $L_{\rm H\beta}$.
Rescaling the strongest predicted optical emission line, [FeXIV]$\lambda$5303, correspondingly
leads to an intensity ratio [FeXIV]$_{\rm{wa}}$/H$\beta_{\rm{obs}} \approx$ 0.01. This compares to the
observed upper limit of [FeXIV]/H$\beta \approxlt 0.1$ as estimated from a spectrum by Peterson et al. (1985)
and it is consistent with a covering of the warm material of less or equal 100\%.

Due to the low emissivity of the warm gas, no strong UV - EUV emission lines are produced
(e.g. HeII\,$\lambda$1640$_{\rm{wa}}$/H$\beta_{\rm{obs}} \le 0.06$,
NeVIII\,$\lambda$774$_{\rm{wa}}$/H$\beta_{\rm{obs}} \le 0.2$, 
FeXVI\,$\lambda$343$_{\rm{wa}}$/H$\beta_{\rm{obs}} \le 2.0$). 
 
Consequently, no known emission line component in NGC 4051
can be fully identified with the warm absorber.
 
This conclusion still holds when the
properties of the warm material are changed compared to the `standard' assumptions
(but note, that only one parameter is varied at a time):   
The strength of e.g. [FeXIV] is also dependent on the density and changes by a factor
of several depending on the value of $n_{\rm H}$, but stays below the observed upper limit.
For subsolar metal abundances of the absorber HeII\,$\lambda$1640 becomes 
the strongest line in the UV-optical 
region due to the increased helium column necessary to ensure the same column in 
metal ions and thereby the same strength of the
absorption edges in X-rays. For metal abundances  
of 0.2 $\times$ solar, we find HeII\,$\lambda$1640$_{\rm{wa}}$/H$\beta_{\rm{obs}} \le 0.35$.  
A strong additional IR spectral component incident on the warm gas slightly 
influences the line emission, although again, there is no strong contribution to the 
optical-UV spectrum.   

\subsubsection {UV absorption lines}

The ionization structure 
of the best-fit warm absorber model indicates that the expected 
UV absorption by e.g. C$^{3+}$ or N$^{4+}$ is low.
A common UV - X-ray absorber has been found in some active galaxies by Mathur
and collaborators (e.g. Mathur 1994).
In the following we give the expected equivalent widths for the UV lines
CIV\,$\lambda$1549, NV\,$\lambda$1240 and Ly$\alpha$ predicted by the warm absorber model.
The column density in C$^{3+}$ is $\log N_{\rm C^{3+}} = 13.45$ which 
yields $N_{\rm C^{3+}}{\lambda}f$ = 10$^{8.09}$ cm$^{-1}$, where $\lambda$ = 1549 $\AA$~and
$f$ = 0.28 is the oscillator strength for CIV (Allen 1955). Performing a standard curve
of growth analysis (Spitzer 1978) this evaluates to give an equivalent width of 
$\log W_{\lambda}$/$\lambda$ = --4.02$^{+0.05}_{-0.08}$ where the uncertainties refer 
to values of the velocity spread parameter $b$ = 100 km/s (for `+') and 20 km/s (for `--')
whereas the central value is calculated with $b$ = 60 km/s. Correspondingly,
we find for NV\,$\lambda$1240 $\log N_{\rm N^{4+}} = 12.90$ 
and $\log 
W_{\lambda}$/$\lambda$ = --4.69
and for Ly$\alpha$
$\log N_{\rm H^{0}} = 15.94$
and $\log W_{\lambda}$/$\lambda$ = --3.04$^{+0.20}_{-0.44}$.              
HST spectra would allow to search for these absorption lines and thereby further constrain the
properties of the warm material. 

\subsection {Time variability of the spectral components and comparison with 
             other observations} 

\subsubsection {Warm absorber}
In the preceding discussion, photoionization equilibrium was assumed. 
That photoionization indeed plays an important role 
for the ionization of warm material is shown by
a direct reaction of the absorber (i.e. the depth of the OVIII absorption edge) 
in MCG-6-30-15 to changes in the continuum (Otani et al. 1996). 
The equilibrium state of the gas depends on 
its reaction timescale compared to the timescale of changes in the continuum flux  
(see Krolik \& Kriss 1995).
In case the continuum variations are slow compared to the recombination timescale,
the warm material re-adjusts to each continuum level, whereas a mean continuum 
is appropriate for modeling in the opposite case. The latter seems to 
apply to the present observation, with 
no reaction of the warm gas despite changes in the luminosity. However, in the long term
there are changes in the ionization parameter of the gas: An earlier \ros 
observation shows both, lower mean luminosity and ionization parameter (McHardy et al. 1995;
treating and re-fitting these data with the same data reduction procedures and model assumptions 
as carried out for the present observation yields
$\log U$ = 0.2, $\log N_{\rm w}$ = 22.45, $\Gamma_{\rm x}$ = --2.2 and 
an integrated (0.1-2.4 keV) luminosity of $L_{\rm x} = 5.4 \times 10^{41}$ erg/s). 
Although this might mean that the warm material 
follows long term trends in the luminosity but not the very short-time variability,
the situation seems to be more complex. McHardy et al. (1995) and  
Guainazzi et al. (1996) find $U$ to be variable within one day. 
McHardy et al. interpret an increase in $U$ in two orbits as a time-delayed 
reaction of the absorber to a continuum high-state. 
The limit on the density of the ionized material estimated by Guainazzi et al.
is $n_{\rm H} \ge 10^7$ cm$^{-3}$. 
(Note an uncertainty in the density estimate of a factor of several resulting from 
e.g. the exact values chosen for the temperature of the warm gas, the ion abundance ratio,
and the estimated time interval.) 
Comparing these observations with the present one, the implications are   
either (i) time gaps just prevented observing a reaction
of the warm gas within the current data, or (ii) the warm material does not see the luminosity changes
in the present observation, 
or (iii) the density of the warm gas varies with time, 
or (iv) the ionization state of the absorber is not dominated by photoionization 
(see Krolik \& Kriss 1995, Reynolds \& Fabian 1995 for some alternatives).  
None of the possibilities can be decided upon with the present data,
but we note the following: 

Possibility (i) would nearly pin down
the density of the warm absorber, to lie in a narrow range around 10$^7$ cm$^{-3}$.  

(ii) A scenario in which
the warm gas does not see luminosity changes is one in which the 
variability is not intrinsic to the central continuum source but caused between
the absorber and the observer. 
Wachter et al. (1988)
proposed the existence of fast moving, dense blobs of matter within the 
inner region of active galaxies, partially blocking the line of sight to the continuum source.
A similar scenario was invoked by Kunieda et al. (1992) to explain different
flux states seen in a \ginga observation of NGC 4051. 
The present observation is consistent with the blob model, if the source were in state `C' 
(in the terminology of Kunieda et al., referring to variable soft and constant hard observed flux). 
We lack simultaneous observations in the harder X-ray region
but placing blobs with a column density of $N_{\rm H} \approx 10^{24.3}$ cm$^{-2}$,
as proposed by Kunieda et al., 
along the line of sight would completely absorb the incident radiation 
in the \ros sensitivity range.  
In that case, no density constraint on the warm material could be derived. 
But the origin of
the blobs and their fast movement (the latter in combination with their large number 
to account for the observed flux states)
remain to be solved. 

(iii) Time-dependent density occurs e.g. if the absorber is in the form of an expanding
cloud or consists of inhomogeneous material e.g. in orbital motion. In the latter case
variability of the observed warm column density is also expected, which indeed is
observed.  
There is approximately a factor of 2 change in warm column between the two \ros observations,
being separated by 2 years (Nov. `91 and Nov. `93), and a factor of 
larger than 10 compared to the \asca data (April `93) 
of Mihara et al. (1994), who derived $\log N_{\rm w} = 21.3 \pm 0.2$.
Guainazzi et al. (1996) find for a second \asca observation (July `94) roughly 
$\log N_{\rm w} \approx$ 22.3, and indications for variability in the OVII edge.     

%
  \begin{figure}  
      \vbox{\psfig{figure=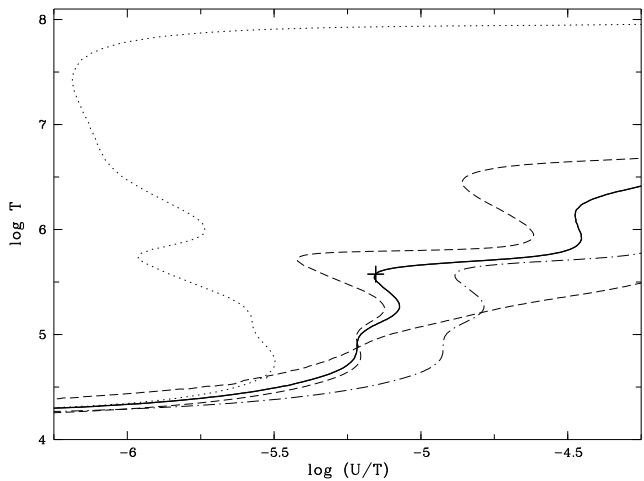,width=8.8cm,%
          bbllx=2.9cm,bblly=1.1cm,bburx=18.1cm,bbury=12.2cm,clip=}}\par
      \vspace{-0.4cm}
 \caption[eq]{Equilibrium gas temperature $T$ against $U/T$
for various SEDs incident on the gas and abundances of the gas. 
The solid line corresponds to the SED of NGC 4051
used for the modeling. The position of the warm absorber is marked. The other 
models correspond to a modification of either metal abundances of 
0.2 $\times$ solar (lower dashed line) or 3 $\times$ solar (upper dashed line)
or spectral shape; dot-dashed: additional EUV black body component
as described in Sect. 3.2, dotted: `mean Seyfert' continuum but with $\Gamma_{\rm x}$ = --1.5,
for comparison.}  
\label{eq}
\end{figure}
In this context, it is also interesting to note that the warm gas in NGC 4051 is 
located near an instable
region in the $U/T - T$ diagram.  
A possible 3 phase equilibrium of the ionized absorber in MCG-6-30-15
was discussed by Reynolds \& Fabian (1995). The results of a similar analysis 
for NGC 4051 are shown in Fig. \ref{eq}.  
Those regions of the equilibrium curve in which the temperature $T$
is multi-valued for constant $U/T$, i.e. pressure, allow for the existence of multiple
phases in pressure balance. 
The parts with negative gradient correspond to thermally unstable equilibria.
For comparison, the corresponding curves for metal
abundances deviating from the solar value are shown.  

\subsubsection {Powerlaw component}    
The possibility of producing all observed {\it spectral} variability in NGC 4051 by warm absorption has been
repeatedly mentioned (e.g. Matsuoka et al. 1990,  
Fiore et al. 1992, Mc Hardy et al. 1995). We find a significantly steeper
powerlaw slope ($\Gamma_{\rm x}$ = --2.3) as compared to other 
observations (e.g. $\Gamma_{\rm x}$ = --1.88, Mihara et al. 1994).  
Consequently, not all soft X-ray spectral variability in NGC 4051 can be traced back to
the influence of the warm absorber.
In particular, the powerlaw slope is also steeper than   
is predicted by currently popular non-thermal pair models (e.g. Svensson 1994).   

Short-timescale variability of $\Gamma_{\rm x}$
during one \ginga observation has been favoured by Matsuoka et al. (1990; 
and has recently been found in \asca data, Guainazzi et al. 1996). 
In the \ginga data, when described by a powerlaw SED, a change of the 2--10 keV
flux by a factor of 3--4 was accompanied by a change $\Delta\Gamma_{\rm{x}}$ = 0.4--0.5.
During the present observation, we find $\Gamma_{\rm x}$ to be essentially constant
despite changes in flux by a factor of larger than 4.
This corroborates the complex and probably time-dependent behaviour of this source.  

\subsection {NLSy1 - character of NGC 4051}

NLSy1 galaxies generally exhibit steep soft X-ray spectra and narrow Balmer lines
(see Boller et al. 1996 for a recent detailed discussion). 
These properties are also shown 
by NGC 4051, in the sense that a simple powerlaw fit to the X-ray spectrum results
in a rather steep powerlaw with $\Gamma_{\rm{x}}$ = --2.9 and the FWHM of H$\beta$ is less than 1000 km/s.
However, much of the X-ray spectral steepness of NGC 4051 is caused by the presence of the warm absorber. 
On the other hand, the intrinsic powerlaw slope is still steeper than the 
canonical one with $\Gamma_{\rm x}$ = --1.9.
And an additional soft excess
is seen in source high-states. This points to           
the complexity of NLSy1 spectra with
probably more than one mechanism at work to cause the X-ray spectral steepness.  

A {\em dusty} environment in NLSy1 galaxies was proposed as one explanation for the narrowness of their
broad lines (Goodrich 1989).   
Dusty {\em warm} gas was suggested to exist in the 
infrared loud quasar IRAS 13349+2438, which has several properties in common with NLSy1 galaxies
(Brandt et al. 1996).  
In case of NGC 4051 no successful description of the X-ray spectrum is achieved when 
compared to models including dust. The general trend holds that it is more difficult to produce 
steep soft X-ray spectra with dusty absorbers. On the contrary, the existence of relatively
stronger absorption edges from more lowly ionized species generally leads to an effective 
{\em flattening} of the spectrum when the edges are not individually resolved.  

\section{Summary and conclusions}

We detect the following spectral components in the Nov. 1993 spectrum of NGC 4051:
A powerlaw in its steepest observed state with $\Gamma_{\rm x}$ = --2.3, 
modified by the presence of a warm absorber, and evidence for a black-body-like soft excess during the
flux high-state with $kT_{\rm bb} \approx$ 0.1 keV. The first two components are 
essentially constant during the observation, but significantly variable 
when compared to former observations. 
Mainly arguments on the number $Q$ of ionizing photons and optical emission line ratios hint to a further 
bump component in the EUV.    

The warm absorber component has been modeled in more detail, which yields 
a column density of $\log N_{\rm w}$ = 22.7 and an ionization parameter of $\log U$ = 0.4
(from X-ray spectral fits), a limit on the density of $n_{\rm{H}} \approxlt$ $3 \times 10^{7}$cm$^{-3}$
(from variability arguments) translating into a 
distance from the nucleus of $r \approxgt 3 \times 10^{16}$ cm, and a covering that can be as large 
as 100 \% (from emission line arguments).   
Observational evidence (no indications of strong reddening along the line of sight, 
e.g. from the UV spectrum) and
model results (no successful X-ray fit) strongly suggest the absorber to be dust free.  

Observable consequences  
of the existence of the ionized material in other spectral regions (in the form of emission or absorption lines) 
are found to be small: None of the observed emission line regions in NGC 4051 can be fully identified with
the warm absorber. The possibility of a contribution to observed individual lines (that would 
complicate line intensity modeling or reverberation mapping), namely to HeII\,$\lambda$1640,
remains in case of subsolar metal abundances of the absorber.   

\begin{acknowledgements}
The \ros project is supported by the German Bundes\-mini\-ste\-rium
f\"ur Bildung und Wissenschaft (BMBW/DARA) and the Max-Planck-Society. 
We are indepted to Gary Ferland for providing {\em{Cloudy}}.
This research has made use of the NASA/IPAC extragalactic database (NED)
which is operated by the Jet Propulsion Laboratory, Caltech,
under contract with the National Aeronautics and Space
Administration.
\end{acknowledgements}

\end{document}